\newcommand{\be}{\begin{equation}}
        \newcommand{\ee}{\end{equation}}
        \newcommand{\ba}{\begin{eqnarray}}
        \newcommand{\ea}{\end{eqnarray}}
        \newcommand{\ban}{\begin{eqnarray*}}
        \newcommand{\ean}{\end{eqnarray*}}
         
\newcommand{\R}{{\mathbb R}}  
  
\newcommand{\Z}{{\mathbb Z}}

\newtheorem{theorem}{Theorem} 
\newtheorem{remark}{Remark} 
 
\newtheorem{corollary}{Corollary} 
\newtheorem{definition}{Definition} 

\documentclass[11pt]{article}
\usepackage {graphicx,latexsym}
\usepackage{amsfonts}

\usepackage[latin1]{inputenc}
\begin{document}

\title{
Curvature function and coarse graining
}
\author{
Homero Díaz-Marín$^{1, 2}$
\footnote{e-mail: \ttfamily homero@matmor.unam.mx} 
and Jos\'e A. Zapata$^1$
\footnote{e-mail: \ttfamily zapata@matmor.unam.mx} 
\\
{\it $^1$Instituto de Matemáticas, Universidad Nacional Autónoma de México} \\ 
{\it A.P. 61-3, Morelia Mich. 58089, México}\\
{\it $^2$Instituto de Física y Matemáticas,}\\ 
{\it Universidad Michoacana de San Nicolás de Hidalgo,}\\
{\it Edif. C-3, C. U., Morelia, Mich.  58040, México}
}

\date{}
\maketitle

\begin{abstract} 
A classic theorem in the theory of connections on principal fiber bundles states that 
the evaluation of all 
holonomy functions gives enough information to characterize the bundle structure 
(among those sharing the same structure group and base manifold) 
and the connection up to a bundle equivalence map. 
This result and other important properties of holonomy functions 
has encouraged their use as the primary ingredient for the construction of families of 
quantum gauge theories. 
However, in these applications often the set of 
holonomy functions used is a discrete proper subset of the set of holonomy functions 
needed for the characterization theorem to hold. 
We show that the evaluation of a discrete set of holonomy functions 
does not characterize the bundle and does not constrain the connection modulo gauge appropriately. 

We exhibit a discrete set of functions of the connection and prove that in the abelian case their 
evaluation characterizes the bundle structure (up to equivalence), and 
constrains the connection modulo gauge up to ``local details'' ignored when working at a given scale. 
The main ingredient is the Lie algebra valued curvature function $F_S (A)$ defined below. It covers the holonomy function in the sense that \\
$\exp{F_S (A)} = {\rm Hol}(l= \partial S , A)$. 
\end{abstract}


\section{Motivation}


We present a reconstruction theorem (Theorem \ref{fReconstr}) 
which characterizes a bundle, modulo bundle equivalence, among the principal fiber bundles with the same base space and structure group $G = U(1)$ using as data the evaluation of a discrete collection of functions of the connection. 
Once the bundle has been identified, the data constrain the connection modulo gauge: 
we show that any connection inducing a given data set can be deformed to a corresponding (singular) connection by means of a local deformation process; see Theorems \ref{AbelianLocalization} and \ref{CflatAsLimits}. 
Our results for non abelian structure groups are not yet as sharp, and they are technically more involved; we will present them in a future publication. 
In order to focus attention to the main issues we restrict several parts of our presentation to base manifolds $M= S^d$; 
the essential ingredient of our construction is of local character, and the extra elements needed to include general base manifolds are well-known. 

The results stated above 
are the main results presented in the article, but we start posing the problem of whether characterization results 
similar to those described above, 
can be reached using as data the evaluation of a set of holonomy functions corresponding to a discrete collection of loops in the manifold. We verify that it is not possible to reconstruct the bundle or to approximately localize the connection modulo gauge. In the remaining of this section we present the motivation in quantum field theory that brought these geometric problems to our attention. The reader who is only interested in the mathematical results, and the reader for whom the physical motivation is obvious can skip the rest of this section. 

The bias of our study was the formulation of effective gauge theories. 
The starting point of an effective theory is an algebra of functions 
that captures the partial knowledge available about the system at a given measuring scale. 
The evaluation of the functions of that algebra is associated with a configuration (or history) of the effective gauge theory of that scale. 
Since much of the system of interest is ignored by the effective description, 
effective theories are based on functions of a coarse graining character. They are supposed to measure certain average properties of the field of interest. A particular type of coarse graining function is one that only depends on the field at certain locations and ignores the rest; these functions are called decimation functions. A clear example of decimation function in gauge theories is a Wilson loop, which uses the holonomy of the connection along a loop as its main ingredient. 

We study the question of whether a discrete collection of holonomy functions is an appropriate collection to be the cornerstone of an effective theory. We will argue that it is not well-suited for that purpose because there are configurations of the system which are macroscopically different but share holonomy data. 

The following simple example captures one of the key ideas behind the structures introduced in this paper. Consider a family of connection 1-forms in two dimensional euclidean space $A_\lambda = \lambda x \wedge {\rm d} y$ and their respective ``magnetic" fields $B_\lambda = {\rm d} A_\lambda$. If we know that the holonomy around one given loop is trivial, we cannot say if the measured field is weak or arbitrarily strong. However, if we 
measure the magnetic flux through a surface that has the loop as its boundary, we can certainly give more information about the strength of the field. In the example situation the holonomy of $A_\lambda$ determines $\lambda$ only up to 
an ambiguity labeled by an integer number, 
whereas, the magnetic flux determines $\lambda$ completely. 

Above we 
exhibited an example 
illustrating the importance of functions of the connection measuring the integral of its curvature on surfaces. 
We will consider families 
of functions of the connection which include 
such curvature functions to gather ``local information about the connection'', 
and we will prove that these families enjoy basic properties that make them good candidates for measuring the connection at a given scale. 
The concrete results are the reconstruction theorem and the approximate localization theorems 
referred to in the first paragraph of this introductory section. 

Curvature evaluations may be of interest to develop effective theories because usually the dynamics of a gauge theory is given by an expression which involves the curvature. The knowledge of functions that measure curvature directly may be advantageous as compared to using holonomy evaluations which are only able to give approximate information about the curvature in the weak field regime. 

With the aim of paving the road to constructing effective theories 
whose configuration space (or space of histories) is the one associated to a discrete family of curvature evaluations, we present a regularization procedure. Each effective configuration can be realized 
as coming from the evaluation on 
a unique connection among a certain family of flat connections with ``conical singularities''. This construction induces a regularization of functions. 

The question of how the mentioned approximation improves as the effective theory is capable of measuring more details of the field is also addressed. 

This paper deals with entirely classical features motivated by a quantization goal. In a separate publication we present a proposal of quantum configuration space (or quantum space of histories) for an effective theory at a given scale 
together with a family of kinematical measures. 
At each scale we have a space which captures information about the connection gathered at this scale, and the physical measure at this scale can be constructed using the kinematical measure and a weighting factor. In that paper also 
a continuum limit is constructed where the expectation values of curvature evaluations 
enjoy of certain independence from the 
auxiliary structures used to define the sequence of effective theories at given scales. 
The construction is compatible with an implementation of Wilson's renormalization group and is related to the loop quantization of gauge theories. 



\section{Holonomy functions in context}
\label{pt3}
\subsection{Geometric and topological properties}
\label{pt3.1}
In theories where the dynamical object is a connection modulo gauge, holonomies 
along based loops can be used as the basic ingredient of a very interesting family of functions. In this subsection we briefly recall their properties.  

Let $(E, \pi , M)$ be a principal fiber bundle and ${\cal A}_\pi$ be the space of connections on it. 
We choose an arbitrary base point $\star \in M$, and consider the space of 
piecewise smooth oriented $\star$-based loops modulo reparametrization. 
We recall that a piecewise smooth curve $\alpha: [ a, b] \to M$ is 
one that can be cut into finitely many pieces $\alpha_i: [ a_i, b_i] \to M$ such that each piece is 
the restriction of a smooth curve $\alpha'_i : ( a_i-\epsilon , b_i+ \epsilon ) \to M$. This condition guarantees regularity; in particular, for this type of curve parallel transport is well-defined. 
One can compose such loops. If we compose one such loop with the loop with opposite orientation we get a {\em thin loop}; if we declare such loops to be null (equivalent  to the constant $\star$-based loop) we get a group 
${\cal L}(\star, M)$. The equivalence relation just introduced is often referred to as retracing equivalence%
\footnote{
For convenience we will write some times $l \in{\cal L}(\star, M)$ and 
some other times $l \in [ l ] \in{\cal L}(\star, M)$. In the first expression we omitted the brackets but we still refer to the equivalence class; only if the context requires a sharper notation we will include the brackets. Also according to the context $l$ will denote a curve modulo reparametrization, or a curve, or the image of the curve. For example we will write $l \subset L \subset M$ meaning that 
the class $[l]$ has a representative whose image 
is contained in the subset $L$ of $M$. 
}. 
Once we have identified $G$ with $\pi^{-1}(\star)$, which can be done by choosing an element of the fiber $b\in \pi^{-1}(\star)$ 
to correspond to ${\rm id} \in G$, 
the holonomy of a given connection $A \in {\cal A}_\pi$ around a loop $l \in {\cal L}(\star, M)$ is a group element 
${\rm Hol}(l , A) \in G$. Moreover, if we fix the connection the holonomy map gives us a group homomorphism 
\[
{\rm Hol}_A \equiv {\rm Hol}( \cdot , A): {\cal L}(\star, M) \to G . 
\]
Gauge transformations act on holonomies in a very simple way because only the evaluation of the gauge transformation on the fiber over the base point plays a role, and the action is by conjugation. 
Holonomy functions can be defined as acting on ${\cal A/G}_{\star , \pi}$, the space of connections modulo the group of gauge transformations whose restriction to the fiber $\pi^{-1}(\star)$ is the identity. In addition, character functions of $G$ can be used to construct gauge invariant functions of the connection which then become functions on ${\cal A/G}_\pi$. 

Another important property of holonomy functions is that they capture information about the curvature of the connection which often plays a central role in the dynamical aspects of a gauge theory. The projected curvature of a connection at a point $p \in M$ can be calculated as a limit of functions of holonomies. We can construct a one parameter family of ``curved parallelograms'' based at $p$ and consider the convergence of certain function of the holonomy in the limit in which the parallelogram becomes infinitesimal (see Section \ref{pt4}). 

The strongest result \cite{Barrett:1991aj} summarizing the geometrical and topological  properties of holonomy evaluations is stated below as a reconstruction theorem which tells us that the family of holonomy evaluations $\{ {\rm Hol}(l , A) \}$ for a fixed connection and all the loops in ${\cal L}(\star, M)$
contains all the information about the connection modulo gauge, and 
even the bundle structure is encoded in it. 

\begin{theorem}[Barrett, Kobayashi] 
\label{ReconstructionContinuum}
Let $M$ be a connected manifold with a base point $\star \in M$, and let 
${\rm H} : {\cal L}(\star, M) \to G$ 
be a group homomorphism which is smooth%
\footnote{
The definition of smoothness  
is axiom H3 in Barrett's article, stating that any smooth finite dimensional family of loops leads to a smooth curve in $G$ after composition with ${\rm H}$. See 
\cite{Barrett:1991aj} for a detailed explanation. 
}. 
Then, there is a differentiable principal fiber bundle $(E, \pi, M, G)$, 
a point $b \in \pi^{-1} (\star)$, and a connection 
$A \in {\cal A}_\pi$ such that ${\rm H} = {\rm Hol}_A $. 
The bundle and the connection are unique up to a bundle equivalence transformation. 
\end{theorem}

\subsection{In an effective theory}
\label{pt3EffTheo}
If we fix a loop, the holonomy ${\rm Hol}( l , \cdot )$ is a function of the connection which is sensitive to the connection only in a very restricted region; thus, holonomies are very natural decimation observables. The strategy of using a discrete collection of holonomy functions as the basic observables of a quantum gauge theory is the basis of lattice gauge theory and loop quantization. As the collection becomes larger (the lattice or graph that hosts the collection of loops becomes finer) the holonomy observables are able to capture more degrees of freedom. 
It may seem natural within these approaches to use one of these sets of observables, a discrete set of holonomy functions determined by a lattice, to construct an effective theory which models the system at a given scale much coarser than the lattice. 
However, even if the lattice is very fine, it would induce a discrete collection of holonomy functions and the reconstruction theorem of the previous subsection may not apply. 

Let $L \subset M$ denote an embedded lattice (or embedded graph). $L$ could be defined as a regular lattice according to a flat metric on $M$, or it could be generated by iterated baricentric subdivisions of a given triangulation, or it could be generated by a random process. Let the base point be a vertex of the lattice, $\star \in L \subset M$, and 
let 
\[
{\cal L}(\star, L) \subset {\cal L}(\star, M)
\]
be the subgroup generated by (classes of) loops (with representatives) whose images are contained in $L$. The holonomy evaluations of a given connection modulo gauge $A \in {\cal A/G}_\pi$ on the loops of ${\cal L}(\star, L)$ form a discrete set with partial 
information about the connection on $\pi$. In fact this set of holonomy evaluations could be characterized by the evaluation on a finite set of generators of ${\cal L}(\star, L)$. 
Now we will study the evaluation of holonomy functions of loops in ${\cal L}(\star, L)$ for all possible smooth connections. We define the $L$-holonomy evaluation map 
\[
{\rm H}_L : {\cal A/G}_{\star , \pi} \to {\cal A/ G}_{\star , L}^{\rm Hol} 
\equiv {\rm hom}({\cal L}(\star, L), G) 
\] 
by ${\rm H}_L(A) = \{ {\rm Hol}( l , A) \}_{l \in {\cal L}(\star, L)}$. 
In our notation 
${\cal A/ G}_{\star , L}^{\rm Hol} \equiv {\rm hom}({\cal L}(\star, L), G)$ 
denotes the group of homomorphisms from ${\cal L}(\star, L)$ to $G$, and 
we call it 
the space of $L$-holonomy evaluations. We do not call it space of $L$-connections because it leads one to assume that it is a good space to host an effective theory of connections at scale $L$. 

Clearly, for every $[A]_L \in {\cal A/ G}_{\star , L}^{\rm Hol}$ the subset 
${\rm H}_L^{-1} ([A]_L) \subset {\cal A/G}_{\star , \pi}$ is uncountable. 
Thus, 
$[A]_L \in {\cal A/ G}_{\star , L}^{\rm Hol}$ captures partial information about the connection modulo gauge; it identifies many distinct configurations 
whose differences will be treated as being ``microscopical in relation to the scale $L$.'' 
Below we will be concerned with the question of whether this partial information and the identifications that it induces are appropriate to build an effective theory of a physical system observed at a given scale. 

What we mean by an effective theory is one whose purpose is to describe the system at a given scale. It should model the behavior of the system as far as the 
measurements available at scale $L$ are concerned. 
In this subsection we analyze the case study 
in which holonomies along the loops of ${\cal L}(\star, L)$ are the available functions. 
The motivation of the work presented in this article is to find a kinematical framework which can host an effective theory at a given scale. 

Perhaps the first issue to study is whether two connections modulo gauge 
$[A] , [A'] \in {\rm H}_L^{-1} ([A]_L) \subset {\cal A/G}_{\star , \pi}$ must be relatively close to each other or not. 
It is simpler to study first the same problem at the non gauge invariant level. For this purpose we will use the notation 
${\rm H}_L : {\cal A}_\pi \to {\cal A/ G}_{\star , L}^{\rm Hol}$. 
In the abelian case we can state the following simple result. 
\begin{theorem}
\label{NoLoc}
Consider a bundle $(E, \pi , M)$ with structure group $G = U(1)$. 
Since ${\cal A}_\pi$ is an affine space two different connections in it define a line. The intersection of the line defined by 
$A , A' \in {\rm H}_L^{-1} ([A]_L) \subset {\cal A}_\pi$ 
and ${\rm H}_L^{-1} ([A]_L) \subset {\cal A}_\pi$ is 
a non compact set which contains the lattice of points of the form 
\[
A_n = A + n(A' - A) \in {\rm H}_L^{-1} ([A]_L) \subset {\cal A}_\pi \mbox{ (for any $n \in \Z$). } 
\]
\end{theorem}
It is important to notice that 
there are gauge invariant functions like the curvature at given base points which are not bounded when evaluated in this set of connections. Thus, connections modulo gauge cannot be (approximately) localized via $L$-holonomy evaluations. 

In the previous section we saw that holonomy evaluations could be used to localize the connection modulo gauge and also to reconstruct the bundle $(E, \pi , M)$ up to bundle equivalence. After we realize that approximate localization fails for $L$-holonomy evaluations, it is natural to study if also the topological information about the bundle is lost when we restrict to a discrete collection of holonomy evaluations. 
Below we show that the bundle structure is not captured by a discrete collection of holonomy evaluations. 
\begin{theorem}
\label{NoRec}
Given any lattice $L \subset S^2$ 
consider the space ${\cal A/ G}_{\star , L}^{\rm Hol}$ of $L$-holonomy evaluations 
of connections in the space ${\cal A/G}_{\star , \pi}$ 
on a $U(1)$ principal fiber bundle $(E, \pi , S^2)$. 
The configuration 
corresponding to trivial holonomy evaluations for all the loops in ${\cal L}(\star, L)$ 
belongs to ${\cal A/ G}_{\star , L}^{\rm Hol}$ 
for every bundle $\pi$. 

Since there are inequivalent $U(1)$ principal bundles over $S^2$ we can conclude that 
a discrete set of holonomy evaluations does not characterize the bundle structure. 
\end{theorem}
{\em Proof.} 
The embedded one dimensional lattice $L\subset S^2$ defines a collection of two dimensional oriented closed surfaces 
$\sigma_k \subset S^2$ whose boundary is a loop in ${\cal L}(\star, L)$ and such that 
$S^2 = \cup \sigma_k$; we will call them {\em plaquettes}. 
Notice that if there was a smooth connection $A \in {\cal A}_\pi$ such that 
$\int_{\sigma_k} dA$ is an integral multiple of $2\pi$ then we would have 
${\rm Hol}( \partial \sigma_k , A) = {\rm id}$. 
We will demonstrate that any given connection 
$A_0 \in {\cal A}_\pi$ can be smoothly deformed to satisfy the curvature condition mentioned above. Before we start, notice that we can order the plaquettes in a sequence in such a way that $\sigma_k$ and $\sigma_{k+1}$ are neighboring plaquettes for every $k$. 
Now we construct the deformation in steps. First, it is easy to see that there is a smooth deformation of the connection which is the identity in the restriction of the connection over 
$S^2 - (\sigma_1 \cup \sigma_{2})$ and such that the new connection 
$A_1 \in {\cal A}_\pi$ satisfies $\int_{\sigma_1} dA_1$ is an integral multiple of $2\pi$. 
The second step does the analogous deformation with support over 
$\sigma_2 \cup \sigma_{3}$ fixing the integral of the curvature in $\sigma_2$ without destroying the job we did over $\sigma_1$. 
The last plaquette automatically satisfies our integral curvature requirement. 
$\Box$ 

In summary, measuring holonomies does not make us aware of the strength of the field. For example, in the case of the magnetic field, 
if we measure the holonomy of the vector potential along a loop and we obtain ${\rm id} \in U(1)$, we cannot say that the magnetic field is close to zero, or even that some regional average of it is dominated by a given bound. 
Additionally, the bundle structure is not captured by holonomy evaluation data. 
If we use holonomies as decimation observables from the continuum, the information that we gather will not tell us the strength of the field and will also fail to characterize the bundle structure up to equivalence. Similarly, if we have a collection of effective gauge theories constructed from a collection of lattices which are ordered by refinement, we cannot decimate from a finer lattice to a coarser one just by evaluating the holonomies on the coarser loops. 
Doing so would result in identifying configurations of the finer lattice which the coarser theory should distinguish.

\subsection{In loop quantization}
The space of generalized connections modulo gauge 
$\overline{\cal A/ G}_{\star , M}$ is defined as a projective limit of spaces 
${\cal A/ G}_{\star , \gamma}^{\rm H}$ \cite{Ashtekar:1994mh} 
(with respect to the coarse graining map given by the pullback of the inclusion map of embedded graphs 
$
\gamma_1 \subset \gamma_2 \Rightarrow {\cal A/ G}_{\star , \gamma_2}^{\rm H} 
\to 
{\cal A/ G}_{\star , \gamma_1}^{\rm H}
$
). 
According to the results of the previous subsection the spaces ${\cal A/ G}_{\star , \gamma}^{\rm H}$ 
are not sensitive to the bundle structure
and cannot localize the connection modulo gauge up to ``microscopic details" in any reasonable sense. 
After the information from all embedded graphs is organized by the projective limit, the issue becomes subtle. 
An important property is that ${\cal A/G}_{\star , \pi}$ can be mapped into 
$\overline{\cal A/ G}_{\star , M}$ injectively 
and its image is a dense subset. This is true for every bundle $(E, \pi , M)$. 

Thus, some generalized connections 
$[\bar{A}] \in \overline{\cal A/ G}_{\star , M}$
contain the information about the bundle 
$(E, \pi , M)$, but if the base manifold admits inequivalent bundles there are other generalized connections 
$[\bar{A}'] \in \overline{\cal A/ G}_{\star , M}$ (a dense subset of them) 
in the same space with the information about the inequivalent bundle structure 
$(E', \pi' , M)$. 

The issue of the localization of the connection modulo gauge once the bundle has been identified is not present in $\overline{\cal A/ G}_{\star , M}$ because a connection modulo gauge is completely determined by the evaluation of its holonomy functions as stated by the reconstruction theorem (Theorem 
\ref{ReconstructionContinuum}). However, we have to keep in mind that the information contained in a single 
${\cal A/ G}_{\star , \gamma}^{\rm H}$ has the deficiencies studied in the previous subsection.

\section{Curvature}
\label{pt4}
%
We saw in the previous section that measuring holonomies does not make us aware of the strength of the field. In this section we recall of another function which is related to the holonomy and does measure the field strength. In the magnetic example of the first section given a surface $S$ bounded by a loop $l = \partial S$ we know that 
$
\exp{i \int_S B} = {\rm Hol}(l= \partial S , A) . 
$
The simple observation is that while ${\rm Hol}(l= \partial S , A)$ does not measure the strength of the field, the magnetic flux does. 
\begin{definition}
\label{curvature}
Consider a $U(1)$ principal fiber bundle $(E, \pi , M)$ and 
a piecewise smooth compact surface with boundary on the base space $S \subset M$. \\
The integral of the curvature ($F = dA$) 
defines a function 
$F_S : {\cal A/G}_{\star , \pi} \to \R$ 
which we call the curvature in $S$, 
\[
F_S(A) = \int_S F . 
\]
\end{definition}
In the non abelian case there is a generalization of this integral. Given a connection on a principal bundle $[A] \in {\cal A/G}_{\star , \pi}$ and a contractible surface with boundary $S \subset M$ such that 
$\star \in \partial S$ we may define 
$F_S([A]) \in {\rm Lie}(G)$ 
when the exponential map 
$\exp : {\rm Lie}(G) \to G$ 
is an onto map. We can lift curves from $G$ to ${\rm Lie}(G)$ unambiguously once we specify a starting point in ${\rm Lie}(G)$. Thus we construct a homotopy $S(t)$ from $\star$ to the surface $S$, and ${\rm Hol}(\partial S(t) , A)$ defines a curve in $G$. The lift of this curve that starts at the origin defines $F_S(A) \in {\rm Lie}(G)$ as its final point. 
The exponential map is not a covering map in general; there is a nonempty subset of $G$ which is the image of points where the differential of $\exp$ is not one to one. We need to restrict the choice of homotopies $S(t)$ to ensure that the resulting curves in $G$ have the same lift. These restrictions are related to the way in which we capture the information about the topology of the bundle. 
In a future publication we will present a study of $F_S$ in the general case. Here we specialize in the abelian case because this simplification lets us give a more complete presentation, and because up to now our results are sharper in the abelian case. 

By construction the function $F_S$ is no more than certain cover of the holonomy function, $\exp{F_S (A)} = {\rm Hol}(l= \partial S , A)$. The only information not present in the holonomy is an integer. 
Due to Stokes theorem if a loop $l$ bounds two compact surfaces with boundary $S, S'$ and there is a three dimensional region with boundary such that 
$\partial R = S \cup (-S')$, then $F_S (A)= F_{S'} (A)$. If the base manifold admits pairs of surfaces with the same boundary which are not related by 
being complementary parts of the boundary of a region, 
the evaluation of the curvature function $F_S$ 
may also give us information about the homology type of $S$. 

If the fundamental group of the base manifold is not trivial many loops are not the boundary of a contractible surface, and the information about their holonomies is not contained in any curvature evaluation. 

The curvature of the connection plays a central role in most classical gauge theories. If we know the curvature functions $F_S$ 
we can write the curvature at a point directly in terms of them. 
Let $S_{\epsilon(v, w)}$ be a ``curved parallelogram'' with one corner at $p\in M$ and whose image in its defining chart is the parallelogram determined by $\epsilon v, \epsilon w \in T_pM$. Then 
\[
F_p(v, w) = \lim_{\epsilon \to 0} \frac{1}{\epsilon^2} F_{S_{\epsilon(v, w)}} . 
\]

\section{Discrete collections of functions of connections}
\label{pt5}
Now we turn to our main objective: to study the properties of discrete collections of functions of connections which include curvature functions. 
Our specific goal is to determine whether some of these collections 
could be used as decimation observables to form the basis of an effective theory. 
We saw in section \ref{pt3.1} that the evaluation of the family of ``all holonomy functions" gave enough information to determine the bundle structure and the 
connection modulo gauge. Similarly, if we evaluate ``all curvature functions" and the holonomy functions corresponding to a discrete set of non contractible loops, we would be able to determine the bundle structure and the connection modulo gauge. 
Now our interest is in a situation where 
the available information regarding the gauge field system is not a complete one; instead, it is decimated by taking a sample: the sample being 
the evaluation of the functions in a discrete subfamily. 

In order to simplify the statement of our results in this section we assume that 
$M = S^{d}$. This topological simplification does not lead to an essentially weaker result because the main issue is of local character. 

First we will 
prove a reconstruction theorem stating that the bundle structure can be recovered from the data generated by evaluation of a  discrete family of curvature functions. 
Once the bundle has been identified, we will proceed to 
study how our data constrain the connection modulo gauge. 
We will show that the connection modulo gauge is determined up to microscopic details in the sense that 
the data determine a unique ``standard" connection which can be constructed by a local deformation process 
of the smooth connection which originated the given data. 

We introduced an embedded one dimensional lattice $L \subset S^d$; now we assume that along with 
the collection of links, 
L also includes a collection of embedded surfaces which could be thought of as lattice plaquettes $\{ \sigma_k \}_L$. The set of surfaces that can be 
made 
with unions of these plaquettes will be denoted by ${\cal S}(L)$. 
The set of conjugacy classes of the group of $L$-loops ${\cal L}(\star, L) \subset {\cal L}(\star, S^d)$ is generated by the classes of loops of the form $\{ l_k = \partial \sigma_k \}$, 
where a path joining the boundary of each plaquette with the base point 
$\star \in L$ is assumed. For the purposes of abelian gauge theories, we can pretend that the group ${\cal L}(\star, L)$ itself is generated by 
$\{  l_k = \partial \sigma_k  \}$. 

Given a connection $A \in {\cal A}_\pi$ the collection of evaluations of the curvature functions assigned to the $L$-plaquettes will be stored in the space $\Omega_{L, \pi}$ and denoted by $\omega_L \in \Omega_{L, \pi}$. 
\[
\omega_L = \{ F_S (A) \}_{S \in {\cal S}(L)} \in \Omega_{L, \pi} . 
\]
By definition $\Omega_{L, \pi} \subset \R^{{\cal S}(L)}$ is the image of the $L$-curvature 
evaluation map from ${\cal A}_\pi$ to $\R^{{\cal S}(L)}$ defined by the set of curvature functions of surfaces in ${\cal S}(L)$. 
Since in the abelian case and inside a local trivialization the curvature functions are simple integrals, 
it is natural to 
see ${\cal S}(L)$ as the group of 2-chains that fit in $L$. Thus, ${\cal S}(L)$ is given the structure of an abelian group and 
the $L$-curvature evaluation map provides a group homomorphism from ${\cal S}(L)$ to $\R$ for each connection $A \in {\cal A}_\pi$, 
$\Omega_{L, \pi} \subset {\rm hom}({\cal S}(L), \R)$. Also, since the curvature form is exact, the curvature function of a boundary vanishes, 
$F_{\partial R} = 0$; it would be sharper to consider $\Omega_{L, \pi}$ 
as homomorphisms from the group homology classes of 2-chains in $L$ to $\R$. 
In addition, there may be obstructions 
due to the bundle structure that further restrict the possible evaluations of curvature functions for connections on 
$A \in {\cal A}_\pi$. For example, given a surface without boundary $F_S(A)$ depends only on the homology class $[S]$ and the bundle $\pi$, and 
it is independent of the particular connection $A \in {\cal A}_\pi$. 

Within the characterization 
$\Omega_{L, \pi} \subset \R^{{\cal S}(L)}$ as the image of the set of curvature functions it is important to notice that ${\cal A}_\pi$ is an affine space and the set of $L$-curvature functions acts as an affine map, which means that $\Omega_{L, \pi} \subset \R^{{\cal S}(L)}$ is also an affine space. 
A more intrinsic characterization of $\Omega_{L, \pi}$ would be desirable. 

The notion of 
independence of a set of curvature functions will have two connotations. The first is when we assume the bundle $\pi$ as known, as we did just above. The second comes from considering 
the connection data as responsible for identifying the bundle. Then our data set
is generated by the evaluation of curvature functions acting on connections of some bundle over $M$ with $U(1)$ structure group, 
\[
\check{\omega}_L = \{ F_S (A) \}_{S \in {\cal S}(L)}  \in \check{\Omega}_L = \cup_\pi \Omega_{L, \pi} \subset \R^{{\cal S}(L)} . 
\]
The subtle difference between $\check{\Omega}_L$ and $\Omega_{L, \pi}$ 
arises because even if the holonomy of a loop 
${\rm Hol}_{\partial \sigma_k} (A)$ is constrained by the knowledge of the holonomy evaluation of other loops, its logarithm $F_{\sigma_k} (A)$ contains information which depends on the bundle structure. This extra data are free in $\check{\Omega}_L$, and is constrained in $\Omega_{L, \pi}$ to be compatible with 
$\pi$. 
We will refer to either $\Omega_{L, \pi}$ or $\check{\Omega}_L$ as the space of $L$-curvature evaluations. 

We remark that the elements of spaces of 
curvature evaluations $\Omega_{L, \pi}$, $\check{\Omega}_L$ 
are determined by the evaluations on the set of lattice plaquettes, $\{ \sigma_k \}_L$, which is a finite set. 
The elements of both sets 
$\Omega_{L, \pi}$ and $\check{\Omega}_L$ contain finite information. 

The relation between the spaces $\check{\Omega}_L$ and ${\cal A/ G}_{\star , L}^{\rm Hol}$ is simple. The exponential map $\R \to U(1)$ 
induces an onto map 
$\exp : \check{\Omega}_L \to {\cal A/ G}_{\star , L}^{\rm Hol}$; according to this map the curvature and holonomy evaluations are related by 
\[
{\rm Hol}_{\partial \sigma_k} (A) = \exp{ (F_{\sigma_k} (A) )} . 
\]

\subsection{Reconstruction}
In this subsection we will 
study bundles with base $M= S^d$ and $G=U(1)$. We will 
prove that given any configuration at scale $L$, 
$\check{\omega}_L \in \check{\Omega}_L$, we can reconstruct the bundle $(E, \pi , M)$ up to bundle equivalence.

We recall that all the principal bundles over a disc are trivial. 
Our base manifold $M = S^d$  can be covered by two charts with the topology of the disc; in fact, we can set $S^d = D^N \cup D^S$ considering the Northern and Southern Hemispheres as closed discs whose intersection is only the equator, 
$D^N \cap D^S = {\rm Eq} \approx S^{d-1}$. 
Thus, the bundle structure is all encoded in the transition functions 
$T^{NS}: S^{d-1} \to G$, which should be thought of as the restriction of smooth transition functions defined in a neighborhood of ${\rm Eq}$ to ${\rm Eq}$. 
A well-known result states that two bundles 
$(E, \pi , M)$, $(E', \pi' , M)$ are equivalent if their corresponding transition functions 
$T^{NS}$ and $T'^{NS}$ are homotopic (for details see for example \cite{Novikov}). 

In the abelian case the homotopy type of the transition functions 
$T^{NS}: S^{d-1} \to U(1)$ can only be non trivial in the case $d=2$ where the type is determined by a winding number. Thus, from now on we focus on the case of a 
$S^2$ base space; all $U(1)$ bundles over $S^d$ for $d \neq 2$ are equivalent. 
Our reconstruction result rests in the fact that the curvature evaluations 
$\check{\omega}_L  \in \check{\Omega}_L$ 
determine the mentioned winding number. 

The local trivializations, 
the embedded lattice, and some other auxiliary structures
satisfy the properties listed below. 
The existence of a local trivialization with these properties is clear. For details see for example \cite{Novikov}. 
\begin{itemize}
\item 
The hemispheres are decomposed as the union of plaquettes 
$D^N= \cup_k \sigma_{k, N}$ and $D^S= \cup_k \sigma_{k, S}$. 
The North Pole and the South Pole are vertices of the embedded lattice. 
\item 
We have chosen a path $\gamma_{SN}$ from the North Pole to the South Pole; this path crosses the equatorial line at $x_0$.  
The path $\gamma_{SN}$ and the point $x_0$ are part of the embedded lattice. 
\item 
For each point in the Equator, $x\in E$ there is a chosen path $\gamma_{x, N}$ from the North Pole to it; there is also an analogous path 
$\gamma_{x, S}$. Among these paths 
only $\gamma_{x_0, N}$ and $\gamma_{x_0, S}$ need to belong to the embedded lattice. 
The one parameter family of curves 
$\gamma_{x, S}^{-1} \circ \gamma_{x, N}$ is a foliation of $S^2 -\{ N, S \}$. 
\item
The construction of the trivializations over the hemispheres is such that the parallel transport along 
all the subpaths of 
$\gamma_{x, N}$ are assigned the identity in the structure group, and a similar condition for the south also holds. 
\item
After the local trivialization over $D^N$ is chosen, the trivialization in the south is chosen so that $T^{NS}(x_0) = {\rm id}$.
\end{itemize}

Due to the above conditions 
\[
T^{NS}(x) = {\rm Hol}
(\gamma_{SN}^{-1} \circ \gamma_{x, S}^{-1} \circ \gamma_{x, N} , A) . 
\]
At scale $L$ we do not have complete knowledge about the transition function $T^{NS}$, but we do know its homotopy type. 
Here is the argument: \\
For each $x \in {\rm Eq}$ the path $
\gamma_{SN}^{-1} \circ \gamma_{x, S}^{-1} \circ \gamma_{x, N}$ is the boundary of a surface, and its holonomy is the exponential of a curvature function. Thus, if we move the point $x$ around the equator towards the east 
the holonomy will trace a curve in $U(1)$ whose lift is a curve in the Lie algebra. We know that this curve starts in $0\in \R$ and ends at $F_{S^2}(A) = \sum_k F_{\sigma_{k, N}}(A) + \sum_k F_{\sigma_{k, S}}(A)$. This information determines the homotopy type of the curve $T^{NS}$ in $U(1)$ and is contained in the data 
$\check{\omega}_L  \in \check{\Omega}_L$. 

Notice that if we have a given embedded lattice we can always choose an equatorial circle, a North Pole, a South and a curve $\gamma_{SN}$ which fit in the lattice. 

We have proved the following reconstruction theorem. 
\begin{theorem}
\label{fReconstr}
Let $(E, \pi , S^d)$ be a smooth principal $U(1)$-fiber bundle with a smooth connection $A$ on it. On the base space $S^d$ we have an auxiliary embedded lattice $L$ 
consisting of a collection of vertices edges and plaquettes 
as defined above. 
Among the $U(1)$-fiber bundles with base space $S^d$, 
the bundle $(E, \pi , S^d)$ 
is characterized (up to bundle equivalence) by the data
\[
\check{\omega}_L = \{ F_S (A) \}_{S \in {\cal S}(L)}  \in \check{\Omega}_L . 
\]
\end{theorem}
After proving that 
$\check{\omega}_L \in \check{\Omega}_L $ 
is enough to characterize the bundle structure the natural question is the following: 
Once the bundle $(E, \pi , S^d)$ has been identified, 
how is that the knowledge of the data 
$\omega_L = \{ F_S (A) \}_{S \in {\cal S}(L)} \in \Omega_{L, \pi}$
constrain the connection? 

We have two arguments motivated by this question. Here is the first one: 
Our partial knowledge can be interpreted as consisting of a collection of local averages of the connection; thus, it would be desirable to have a result stating that two connections which have the same local averages can be related by means of a local deformation which washes away their microscopical differences while preserving the given macroscopical averages. 
In the following section 
(Corollary \ref{CurvEvalDetermsDeltaFlat}) 
we prove that the data $\omega_L \in \Omega_{L, \pi}$ determine a unique (singular) connection of a certain ``standard type", and 
that any 
smooth connection modulo gauge 
$[A] \in {\cal A/G}_{\star , \pi}$ producing the data $\omega_L(A) = \omega_L$ 
can be taken to the standard connection determined by $\omega_L$ by means of local deformations (Theorem \ref{CflatAsLimits}). 

Our second argument is stated in the theorem below. 
Recall that in the abelian case $\Omega_{L, \pi}$ is an affine space, and that the $L$-curvature evaluation map 
\[
\omega_L : {\cal A}_\pi \to \Omega_{L, \pi} , 
\]
defined by 
$\omega_L(A) = 
\omega_L = \{ F_S (A) \}_{S \in {\cal S}(L)} \in \Omega_{L, \pi}$, 
is an affine map. 
Then the following result is immediate. 
\begin{theorem}
\label{AbelianLocalization}
Let $(E, \pi , S^d)$ be a smooth principal $U(1)$-fiber bundle, and 
${\cal A}_\pi $ be the space of connections on it. 
Given any convex neighborhood $B \subset \Omega_{L, \pi}$ 
\[
\omega_L^{-1} (B) \subset {\cal A}_\pi \mbox{ is convex .}
\]
\end{theorem}
To appreciate the value of this property of $L$-curvature evaluations is enough to see that it is not present when only holonomies are used to gather information about the connection; compare with the statement in Theorem \ref{NoLoc}.

\section{Regularization}
\label{pt6}
We have seen that the $L$-curvature evaluation map 
\[
{\cal A/G}_{\star , \pi} \to \check{\Omega}_L
\]
is many to one. The reason is double; one possible reason is that there may be global information that cannot be captured by curvature functions. This happens in base manifolds that have non contractible loops where the holonomy along these loops cannot be captured by the curvature function of a surface. In order to avoid this possibility, we will momentarily restrict to $M = S^d$. The second reason that makes the above map many to one is that local differences on a pair of connections may not be noticeable given the resolution of the discrete set of $L$-curvature functions. 

In this section we will be concerned with defining ``regularity conditions'' that 
can be assumed with the intention of 
finding a restricted set of configurations within the space of configurations of the continuum 
such that one and only one configuration corresponds to the evaluation map of the $L$-curvature functions. 
Then we will be able to consider the space of evaluations of the $L$-curvature functions 
$\check{\Omega}_L$ as the space of configurations (or histories) of an effective theory and regularize some functions of smooth connections by the pull back of the map 
$\check{\Omega}_L \to \mbox{``Smooth connections''}$ defined by our ``regularity conditions''. A detailed treatment will be given below. 
We will use a set of regularity conditions that we called $C$-flat in an earlier work 
\cite{Martinez:2005pj}; 
it needs 
that the lattice $L$ be of a certain type. The structure needed will be that of a cellular decomposition of the manifold and the regularity conditions state that the connection be flat 
when restricted to the interior of each of the cells, 
allowing (distributional) curvature only on the $d-2$ 
skeleton (which is formed by the union of the closed $d-2$ dimensional cells). 
There are other regularization frameworks, for example one can use a triangulation and 
a regularity condition stating that 
the curvature be homogeneous inside the $d$ dimensional simplices according to the euclidean metric that comes with them. 

In order to simplify our presentation the cellular decompositions that we will use are simplicial decompositions (triangulations). 
To maintain clarity in the following definition we will summarize our notation for simplicial decompositions of a smooth manifold $M$, 
as well as 
for the operation of baricentric subdivision, 
for the notion of the $n$-skeleton of a simplicial complex and for the simplicial representation of a curve and that of a surface. 
For a detailed account of the subject see for example 
\cite{fritsch-piccinini, FoundationalEssays}. 
\begin{definition}
\label{triangulations}
A {\bf simplicial decomposition} $( | \Delta | , \phi )$ of a smooth manifold $M$ 
is composed by a simplicial manifold $| \Delta |$ and a compatible 
homeomorphism $\phi : | \Delta | \to M$. The compatibility condition 
asks that the embedded simplices be smooth and it 
will be described below. In the literature this structure is called a smooth triangulation or a Whitehead triangulation. It is known that every smooth manifold admits smooth triangulations and that any two such triangulations are equivalent in the sense that the simplicial manifolds $|\Delta | , |\Delta' |$ are related by a piecewise linear map 
\cite{FoundationalEssays}. 

We can see 
$( | \Delta | , \phi )$ as a collection of simplices of dimensions from zero to $d$ embedded in $M$, which cover $M$  
in such a way that two embedded simplices can either be disjoint or intersect in another embedded simplex of the collection. 

The simplicial manifold $| \Delta |$ carries an euclidean structure in each of its simplices; 
the compatibility condition states that 
the restriction of $\phi$ to each of the simplices be a diffeomorphism to its image in $M$ (the restriction of a diffeomorphism of an open neighborhood of the simplex as a subset of the euclidean space where $| \Delta |$ is contained on to a neighborhood of the image of the simplex in $M$). 

The {\bf baricentric subdivision} ${\rm Sd}|\Delta |$ of $| \Delta |$ 
is a simplicial manifold that is identified with $|\Delta|$, but which carries a finer triangulation. It 
has one vertex for each simplex of $| \Delta |$ which is located at the baricenter of the simplex according to its euclidean structure. A subset of $n$ such vertices defines a $n-1$ dimensional simplex of ${\rm Sd}|\Delta |$ 
if the corresponding subset of simplices of $| \Delta |$ can be ordered by inclusion; 
in this case, it defines a geometric simplex contained in the higher dimensional simplex of the set. 
The baricentric subdivision can be used to define 
a simplicial decomposition which will be called $({\rm Sd}|\Delta |, \phi )$. 

The subset of $n$ dimensional simplices of $| \Delta |$ will be denoted by $| \Delta |^n$. The {\bf $n$ skeleton} of $| \Delta |$ is 
$|\Delta |^{(n)} = \cup_{m=0}^n | \Delta |^m$; if we need to refer to the embedded simplicial complex we will write $\phi (|\Delta |^{(n)})$. 
We will write $( | \Delta |^n , \phi )$ or $( | \Delta |^{(n)} , \phi )$ when we refer to either 
subset of embedded simplices of $( | \Delta | , \phi )$. 

The {\bf simplicial representation} of a piecewise smooth curve $c \subset M$ (if it exists) 
is the only simplicial curve $c_1 \subset \phi ({\rm Sd}|\Delta |^{(1)})$ whose intersection type with $( | \Delta | , \phi )$ is of the same type as that of $c$. A simplicial representation would not exist if the intersection type of $c$ and $( | \Delta | , \phi )$ cannot be matched by that of a simplicial curve due to its required finiteness. 
The simplicial representation can 
be constructed as follows: First divide $c$ into connected components $c_\tau$ whose interior intersects the interior of only one simplex $\tau \in ( | \Delta | , \phi )$. The segment can intersect many simplices, but 
${\rm Int}(c_\tau) \cap {\rm Int}(\tau') \neq \emptyset \Rightarrow \tau = \tau'$. 
The simplicial representative of $c_\tau$ is the simplicial curve 
$c_{\tau, 1}\subset \phi ({\rm Sd}|\Delta |^{(1)})$ whose vertices are 
$\{ \sigma \in ( | \Delta | , \phi ) \mbox{ such that } 
c_\tau \cap {\rm Int}(\sigma) \neq \emptyset \}$, where we consider that the interior of a 
simplex is the simplex minus its boundary, and 
our definition needs that the interior of a 
zero dimensional simplex (a vertex) be considered as the vertex itself. 
The curve $c_1$ does not carry a parametrization; it is (the image by $\phi$ of) a simplicial curve: a collection of neighboring links and vertices. If $c$ has an orientation it is clear that $c_1$ inherits it. 

Similarly, the simplicial representation of a piecewise smooth surface $S \subset M$ (if it exists) is the only simplicial surface 
$S_1 \subset \phi ({\rm Sd}|\Delta |^{(2)})$ 
whose intersection type with $( | \Delta | , \phi )$ is of the same type as that of $S$, and it can be constructed using the location of its vertices as done above. 
Also, an orientation in $S$ would induce one in $S_1$. 
\end{definition}

From now on we will work with curves and surfaces that have simplicial representations. In particular, we require that the elements of 
${\cal L}(\star, M)$ are classes of loops with at least one representative that has a simplicial representation. 
For this reason we will restrict our treatment to the subgroup of 
piecewise linear loops  
${\cal L}(\star, M)_{\rm PL} \subset {\cal L}(\star, M)$ where the PL structure in $M$ is the one induced by the triangulation $( | \Delta | , \phi )$. Also surfaces will be required to be PL. We will denote the set of 
oriented closed PL surfaces by 
${\cal S}(M)_{\rm PL}\subset {\cal S}(M)$. 
We recall that a PL curve (surface) in $M$ is a piecewise smooth curve (surface) 
whose intersection with each of the simplices of $( | \Delta | , \phi )$ is 
either empty, or it is 
a piecewise affine curve (surface) according to the euclidean structure inherited from the simplex in $|\Delta|$. Thus, $[l] \in{\cal L}(\star, M)$ is considered to be PL, 
$[l] \in{\cal L}(\star, M)_{\rm PL}$, if it has a representative $l \in [l]$ that is a piecewise linear curve. 

Our embedded one dimensional lattice $L$ which defines the group 
 ${\cal L}(\star, L)$ is taken to be $L = \phi ({\rm Sd}|\Delta |^{(1)})$. 
 We had asked that the base point be inside the lattice $\star \in L \subset M$. For reasons that we will give in the next section, we need that the base point be the baricenter of one of the maximal dimension simplices of $( | \Delta | , \phi )$, 
 $\star \in \phi ( |\Delta|^d \subset {\rm Sd}|\Delta |^0)$. 
The discrete family of curvature evaluations 
that we consider at scale $L$ is the one 
induced by the family of surfaces 
whose image fits in 
$\phi({\rm Sd}|\Delta |^{(2)})$. The set of such surfaces will be denoted by 
${\cal S}(L) \subset {\cal S}(M)_{\rm PL}$. 
The evaluation of such curvature functions is determined by the evaluation of the curvature in the set of ``plaquettes'', the embedded two simplices of 
$({\rm Sd}|\Delta |^2, \phi )$. 

The (classes of) loops in 
 ${\cal L}(\star, M)_{\rm PL}$ will be divided into equivalence classes 
according to the way in which they intersect the simplices of 
$(| \Delta | , \phi )$. 
\begin{definition}
\label{ClassesOfLoops}
Given a loop in a class $l \in [l] \in{\cal L}(\star, M)_{\rm PL}$ 
consider the class of its simplicial representation $[ l_1 ] \in {\cal L}(\star, L)$. 
It is easy to see that $[ l_1] \in {\cal L}(\star, L)$ 
is independent of the choice of representative $l \in [l]$. Thus, we 
will write $[l]_1$ instead of $[ l_1]$, and we have an assignment 
\[
{\cal L}(\star, M)_{\rm PL} \to {\cal L}(\star, L) 
\]
which divides ${\cal L}(\star, M)_{\rm PL}$ into equivalence classes. 
Similarly, for surfaces we have 
\[
{\cal S}(M)_{\rm PL} \to {\cal S}(L) 
\]
which divides ${\cal S}(M)_{\rm PL}$ into equivalence classes. 
\end{definition}

Now we will define $\Delta$-flat connections.  
A connection induces a holonomy map from ${\cal L}(\star, M)_{\rm PL}$ to $G$ and a curvature map from ${\cal S}(M)_{\rm PL}$ to $Lie(G)$. 
In the case of a $\Delta$-flat connection the evaluation of holonomy and curvature maps 
depends only on the equivalence class of the loop or surface.
This requirement implies that nontrivial $\Delta$-flat connections are not smooth (or even continuous) as $\pi$ connections. 
%
We will relax the 
smoothness and continuity requirements 
allowing curvature singularities, while retaining the property that each $\Delta$-flat connection captures the topology of $\pi$. 
A $\Delta$-flat connection can be thought as a distribution constructed as a limit of a sequence of $\pi$ connections. 
\begin{definition}
\label{DeltaSmooth}
\begin{itemize}
\item 
${\cal A}_{\pi_\tau}^\infty$ is the space of smooth connections 
in the bundle $\pi_\tau = \pi |_{\tau \in (|\Delta | , \phi)}$ (restrictions to $\pi_\tau$ 
of connections which are smooth in an open neighborhood of the subbundle $\pi_\tau$ of $\pi$). 
\item 
$\bar{\times}_{\tau \in (|\Delta | , \phi)}  {\cal A}_{\pi_\tau}^\infty$ 
is the subset of the cartesian product 
defined by the compatibility condition 
$\sigma \subset \tau \Rightarrow A_\sigma = A_\tau |_\sigma$. 
We will write \\
${\cal A}_\pi^{\Delta\mbox{-}\infty} = 
\bar{\times}_{\tau \in (|\Delta | , \phi)}  {\cal A}_{\pi_\tau}^\infty$. 
\item 
A $\Delta$-smooth bundle map is a 
homeomorphism of the total space $\tilde{f}: E \to E$ which sends fibers to fibers, and 
such that for each simplex $\tau \in (|\Delta | , \phi)$ the set $\pi^{-1}(\tau)$ is 
preserved, and $\tilde{f}|_{\pi^{-1}(\tau)}$ is a diffeomorphism. It induces a homeomorphism $f: M \to M$ in the base space which preserves each simplex and whose restriction to each simplex is a diffeomorphism.  Clearly a $\Delta$-smooth bundle map acts on connections by pull back and sends 
${\cal A}_\pi^{\Delta\mbox{-}\infty}$ to itself. 
\item 
${\cal G}_{\star, \pi}^{\Delta\mbox{-}\infty}$ 
is the group of $\Delta$-smooth bundle equivalence maps which consists of 
$\Delta$-smooth bundle maps which induce the identity map on the base space $M$. 
We will write 
${\cal A/G}_{\star , \pi}^{\Delta\mbox{-}\infty} = 
{\cal A}_\pi^{\Delta\mbox{-}\infty} / {\cal G}_{\star, \pi}^{\Delta\mbox{-}\infty}$. 

\end{itemize}
\end{definition}

Holonomy and curvature functions determined by loops and surfaces in 
${\cal L}(\star, M)_{\rm PL}$ and ${\cal S}(M)_{\rm PL}$ 
respectively are defined in 
${\cal A/G}_{\star , \pi}^{\Delta\mbox{-}\infty}$. 
$\Delta$-flat connections modulo gauge will be characterized by the value that holonomy and curvature functions have when evaluated on them. Since we will relax continuity, the knowledge of all the holonomy evaluations will not determine uniquely the evaluation of the curvature functions. 
The set of values that holonomy and curvature functions can obtain when evaluated on 
a $\Delta$-flat connection will be restricted by the bundle $\pi$; this is how each $\Delta$-flat connection stores the information about the bundle structure. %
\begin{definition}
\label{DeltaFlat}
A $\Delta$-flat connection modulo gauge 
\[
[A_\Delta] \in {\cal A/G}_{\star , \pi}^{\Delta\mbox{-flat}} 
\]
is characterized by the set of its holonomy and curvature evaluations 
for every loop $l \in {\cal L}(\star, M)_{\rm PL}$ and 
for every surface $S \in {\cal S}(M)_{\rm PL}$. These evaluations must satisfy 
the following requirements: 
\begin{enumerate}
\item 
There is a sequence 
$[A_n] \in {\cal A/G}_{\star , \pi}^{\Delta\mbox{-}\infty}$ such that 
\[
{\rm Hol}_l(A_\Delta) = \lim_{n \to \infty} {\rm Hol}_{l}(A_n)  \quad \quad , \quad \quad 
F_S(A_\Delta) = \lim_{n \to \infty} F_{S}(A_n) . 
\]
\item 
$$
F_S(A_\Delta) = F_{S_1}(A_\Delta) 
$$
where $S_1$ denotes the simplicial representation of the surface $S$ 
(see definition \ref{triangulations}). 
\end{enumerate}
\end{definition}
We remark that the set of evaluations of holonomy and curvature functions on 
$\Delta$-flat connections inherits the property 
$\exp{F_S (A_\Delta)} = {\rm Hol}_{\partial S}(A_\Delta)$, and that a direct consequence of the definition is  
${\rm Hol}_l(A_\Delta) = {\rm Hol}_{l_1}(A_\Delta)$ for every loop 
$l \in {\cal L}(\star, M)_{\rm PL}$.

\begin{theorem}
\label{CflatAsLimits}
To every smooth connection modulo gauge $[A_0] \in {\cal A/G}_{\star , \pi}$ there corresponds a unique $\Delta$-flat connection modulo gauge 
$[A_\Delta] \in {\cal A/G}_{\star , \pi}^{\Delta\mbox{-flat}}$ such that 
for every $l \in {\cal L}(\star, L)$ and every $S \in {\cal S}(L)$  
\[
{\rm Hol}_l(A_\Delta) = {\rm Hol}_l(A_0) 
\, \, \, \, {\mbox and } \, \, \, \, F_S (A_\Delta) = F_S (A_0)  . 
\]
This $\Delta$-flat connection can be constructed as a local deformation of 
$[A_0] \in {\cal A/G}_{\star , \pi}$ in the sense that there is a sequence of 
$\Delta$-smooth bundle maps (see definition \ref{DeltaSmooth}) which produce a sequence of connections which converge to 
$[A_\Delta] \in {\cal A/G}_{\star , \pi}^{\Delta\mbox{-flat}}$. 
\end{theorem}
The proof will be given at the end of this section. 

\begin{corollary}
\label{CurvEvalDetermsDeltaFlat}
In the case $M=S^d$ 
a point in the space of curvature evaluations 
\[
\check{\omega} \in \check{\Omega}_L
\]
determines a unique 
$\Delta$-flat connection modulo gauge 
$[A_\Delta] (\check{\omega}) \in {\cal A/G}_{\star , \pi(\check{\omega})}^{\Delta\mbox{-flat}}$. 
\end{corollary}
{\em Proof.}\\
By definition of the space $\check{\Omega}_L$, for every $\check{\omega} \in \check{\Omega}_L$ 
there is a bundle $\pi$ over $S^d$ and 
$[A_0] \in {\cal A/G}_{\star , \pi}$ such that all the $L$-curvature evaluations of $[A_0]$ 
agree with the data in $\check{\omega}$. 
The characterization theorem (Theorem \ref{fReconstr}) tell us that the data in $\check{\omega}$ characterize $\pi(\check{\omega})$ up to bundle equivalence. Thus, even when there can be many connections compatible with the data $\check{\omega}$, all of them essentially live in the same bundle. 

The previous theorem implies that there is a unique $\Delta$-flat connection $[A_\Delta]$ which shares the data generated by 
$L$-holonomy evaluations and $L$-curvature evaluations with $[A_0] \in {\cal A/G}_{\star , \pi}$. 
For the purposes of abelian gauge theories, in the case 
$M=S^d$, 
$\{ l_k = \partial \sigma_k \}_{k \in N}$ 
can be treated as a set of generators of 
${\cal L}(\star, L)$. Thus, $L$-curvature evaluations determine 
all the $L$-holonomy evaluations. Then, the $\Delta$-flat connection $[A_\Delta]$ is determined by the 
$L$-curvature evaluations (the data in $\check{\omega}$) alone. 
Since any other choice of smooth connection $[A'_0] \in {\cal A/G}_{\star , \pi(\check{\omega})}$ would necessarily share the data of 
$L$-curvature evaluations (and then $L$-holonomy evaluations), the constructed $\Delta$-flat connection $[A_\Delta]$ is uniquely determined by 
$\check{\omega}$. 
$\Box$ 

Now we return to our goal of regularizing functions of smooth connections. 
In order to give more elements towards the delicate issue of regularization we present two complementary perspectives. First, it is natural to regularize functions using the sequences that define $\Delta$-flat connections. If $M = S^d$ and we assume the bundle as given, there is a map 
$\Omega_{L, \pi} \to {\cal A/G}_{\star , \pi}^{\Delta\mbox{-flat}}$. 
Thus, given a function 
$H: {\cal A/G}_{\star , \pi}^{\Delta\mbox{-}\infty} \to \R$ we may define 
\[
H^{\rm reg}(\omega) = \lim_{n \to \infty} H(A_n)
\]
where the sequence 
$[A_n] \in {\cal A/G}_{\star , \pi}^{\Delta\mbox{-}\infty}$ defines the $\Delta$-flat connection modulo gauge 
$[A_\Delta](\omega)$ (see definition \ref{DeltaFlat}). For the regularization to exist the limit should exist and be independent of the defining sequence. 

The second regularization strategy is to find a space of smooth connections in which we can inject the space of $\Delta$-flat connections and define the regularized functions by pull back. 
The bundle $(E, \pi , M)$ can be pulled back to $M - \phi (| \Delta |^{(d-2)})$; we call this bundle $\pi^-$. 
Consider 
${\cal G}_{\star, \pi^-(\pi)} \subset {\cal G}_{\star, \pi^-}$ consisting of 
of bundle equivalence maps of $\pi^-$ which can be extended to $\pi$ resulting in bundle maps which are smooth everywhere 
except possibly on $\pi^{-1} (\phi (| \Delta |^{(d-2)}))$ where they are continuous. 
By construction 
\[
({\cal A/G}_{\star, \pi} )|_{\pi^-} \subset {\cal A}_{\pi^-} / {\cal G}_{\star, \pi^-(\pi)} . 
\]
Given a surface $S \in {\cal S}(M)_{\rm PL}$ such that 
$\partial S \cap \phi (| \Delta |^{(d-2)}) = \emptyset$, define 
$F_S(A) = \int_{\partial S} A$.%
\footnote{
If needed we consider 
$\int_{\partial S} = \int_{l_1} + \ldots + \int_{l_m}$ with each $l_i$ contained in a trivializing open set. 
} 
We know that even when ${\rm Hol}_{l= \partial S}$ is ${\cal G}_{\star, \pi^-}$ 
invariant; $F_S(A)$ is not. The origin of this phenomenon is the 
presence of gauge transformations whose restriction to 
the connected components of 
$\partial S$ induce non contractible curves in $U(1)$. These gauge transformations do not belong to ${\cal G}_{\star, \pi^-(\pi)}$; it is easy to verify that 
$F_S(A)$ is a well-defined function in 
${\cal A}_{\pi^-} / {\cal G}_{\star, \pi^-(\pi)}$. 
For the same reasons, 
we can write 
$
({\cal A/G}_{\star , \pi}^{\Delta\mbox{-}\infty} )|_{\pi^-} \subset 
{\cal A}^{\Delta\mbox{-}\infty}_{\pi^-} / {\cal G}^{\Delta\mbox{-}\infty}_{\star, \pi^-(\pi)}
$, 
where
${\cal A}_{\pi^-}^{\Delta\mbox{-}\infty} = 
\bar{\times}_{\tau \in (|\Delta | , \phi)}  {\cal A}_{\pi_{\tau^-}}^\infty$, and 
$\pi_{\tau^-}$ is the pullback of $\pi$ to $\tau^-= \tau - \phi (| \Delta |^{(d-2)}) \subset M$. 
Similarly,  for surfaces $S \in {\cal S}(M)_{\rm PL}$ such that 
$\partial S \cap \phi (| \Delta |^{(d-2)}) = \emptyset$, the functions
$F_S$ are defined in 
$
{\cal A}^{\Delta\mbox{-}\infty}_{\pi^-} / {\cal G}^{\Delta\mbox{-}\infty}_{\star, \pi^-(\pi)}
$. 

\begin{definition}
${\cal A/G}^{\Delta\mbox{-}\infty}_{\star, \pi^-(\pi)}$ 
is the subset of 
$
{\cal A}^{\Delta\mbox{-}\infty}_{\pi^-} / {\cal G}^{\Delta\mbox{-}\infty}_{\star, \pi^-(\pi)}
$
which contains $({\cal A/G}_{\star , \pi}^{\Delta\mbox{-}\infty} )|_{\pi^-}$ and 
also the connections modulo gauge 
$
[A] \in {\cal A}^{\Delta\mbox{-}\infty}_{\pi^-} / {\cal G}^{\Delta\mbox{-}\infty}_{\star, \pi^-(\pi)}
$ such that there is a sequence 
$[A_n] \in ({\cal A/G}_{\star , \pi}^{\Delta\mbox{-}\infty} )|_{\pi^-}$ satisfying 
\[
\lim_{n \to \infty} {\rm Hol}_l(A_n) = {\rm Hol}_l(A) \quad \quad , \quad \quad 
\lim_{n \to \infty} F_S(A_n) = F_S(A)
\]
for each 
$l \in {\cal L}(\star, M - \phi (| \Delta |^{(d-2)}))_{\rm PL}$, and 
each $S \in {\cal S}(M)_{\rm PL}$ with 
$\partial S \cap \phi (| \Delta |^{(d-2)}) = \emptyset$. 

The topology in ${\cal A/G}^{\Delta\mbox{-}\infty}_{\star, \pi^-(\pi)}$ is 
induced from 
$
{\cal A}^{\Delta\mbox{-}\infty}_{\pi^-} / {\cal G}^{\Delta\mbox{-}\infty}_{\star, \pi^-(\pi)}
$. 
\end{definition}
From the definition of $\Delta$-flat connections it is clear that 
\[
({\cal A/G}_{\star , \pi}^{\Delta\mbox{-flat}})|_{\pi^-} \subset  
{\cal A/G}^{\Delta\mbox{-}\infty}_{\star, \pi^-(\pi)} . 
\]
Regularization of functions from the continuum is given by the pull back of the 
restriction 
map 
whenever the functions have a well-defined extension (by continuity in 
${\cal A/G}_{\star, \pi^-(\pi)}^{\Delta\mbox{-}\infty}$) 
to act on connections with conical singularities. 

A significant family of examples is the regularization of curvature and holonomy functions for surfaces and loops which do not necessarily fit in 
${\cal S}(L) \subset {\cal S}(M)_{\rm PL}$ or 
${\cal L}(\star, L) \subset {\cal L}(\star, M)_{\rm PL}$ respectively. 
\begin{corollary}
\label{regF}
Given any $S \in {\cal S}(M)_{\rm PL}$ the function 
$F^{\rm reg}_S : \check{\Omega}_L \to \R$ 
is well-defined. 
If $\omega = \omega(A) \in \Omega_{L, \pi(\check{\omega})} \subset \check{\Omega}_L$, 
for a smooth connection modulo gauge 
$[A] \in {\cal A/G}_{\star , \pi(\check{\omega})}$
then 
\[
F^{\rm reg}_S(\omega(A)) = F_{S_1}(A) . 
\]
The regularized curvature function of a surface $S \cup S' \in {\cal S}(M)_{\rm PL}$, where 
$S \cap S'$ is contained in the boundary of each of the surfaces $S$, $S'$, satisfies 
\[
F^{\rm reg}_{S \cup S'} = F^{\rm reg}_S + F^{\rm reg}_{S'} . 
\]

Since the base space considered is $M=S^d$, we can regularize holonomy functions to act on 
the space of curvature evaluations. Given any $l \in {\cal L}(\star, M)_{\rm PL}$, 
and any surface $S \in {\cal S}(M)_{\rm PL}$ which has $l$ as boundary, 
the function 
${\rm Hol}^{\rm reg}_l \doteq \exp{(F^{\rm reg}_S)} : \check{\Omega}_L \to \R$ is well-defined. 
If $\omega = \omega(A) \in \Omega_{L, \pi(\check{\omega})} \subset \check{\Omega}_L$, 
for a smooth connection modulo gauge 
$[A] \in {\cal A/G}_{\star , \pi(\check{\omega})}$
then 
\[
{\rm Hol}^{\rm reg}_l(\omega(A)) = {\rm Hol}_{l_1}(A) . 
\]
\end{corollary}
{\em Proof.} 
The part on the existence and evaluation of the regularized curvature and holonomy functions is a direct consequence of the two previous theorems. The properties of such evaluations follow from properties of the simplicial representation of surfaces and curves 
\begin{itemize}
\item 
Given two surfaces such that $S \cap S' \subset \partial S \cap \partial S'$ we can consider $S \cap S' \in {\cal S}(M)_{\rm PL}$; its simplicial representation satisfies
\[
(S \cap S' )_1 = S_1 \cap S'_1 .
\]
\item 
Given any $S \in {\cal S}(M)_{\rm PL}$ we have
\[
(\partial S)_1 = \partial S_1  .
\]
\end{itemize}
These properties are equalities among subsets of 
$\phi({\rm Sd}|\Delta |^{(2)})$ and $\phi({\rm Sd}|\Delta |^{(1)})$
which can be verified directly from the definition of simplicial representative. $\Box$

Other examples of functions that can be regularized are functions which can be written in terms of the curvature functions that we just studied. Among those examples is 
a regularization of the Euler character in the case of a two dimensional compact base manifold 
\[
e(A) = \sum_{\tilde{\sigma} \in 
(({\rm Sd}|\tilde{\Delta} |)^2 , \tilde{\phi})} 
\int_{\tilde{\sigma}} dA \quad 
\mapsto  \quad 
e^{\rm reg} (\check{\omega}) = 
\sum_{\tilde{\sigma}} 
F^{\rm reg}_{\tilde{\sigma}}(\check{\omega}) , 
\]
where $(({\rm Sd}|\tilde{\Delta} |) , \tilde{\phi})$ is any triangulation of $M$ with piecewise linear simplices according to the PL structure that we are using. 
The function $e: {\cal A/G}_{\star, \pi} \to \R$ turns out to 
depend only on the bundle and not in the specific connection over it.
It is regularized to act on $\check{\Omega}_L$ as a sum of non trivial functions, but 
our previous results imply that 
$e^{\rm reg} (\check{\omega}(A)) = e(A)$ 
which depends on $\pi(\check{\omega})$ and not on the specific smooth connection on it. 
If the base space has higher dimension than two, for any embedded surface we can define 
$e_S^{\rm reg} (\check{\omega})$, and obtain similar results. 

Two dimensional gravity could be an interesting example within reach. We would have to extend the framework 
to include a frame field and define dynamics as a regularization of the first order Einstein action a la Palatini, which may lead us to 
the first order Regge action \cite{1stRegge} in two dimensions. Another route would be to 
construct general relativity as a constrained $SO(2)$-$BF$ theory. 
A first treatment in these cases may be simple since the $B$ field is integrated out leading to the restriction $F=0$ which is naturally implemented in terms of the curvature functions defined in this paper. We would need to compare with the work presented in \cite{Oriti:2004qk}. 

Our regularization procedure is also available for non abelian connections; $C$-flat connections were introduced in \cite{Martinez:2005pj} without a restriction asking the group to be abelian. The refinements of the notion of $C$-flat connections contained in this paper are naturally extended to non abelian cases; they rest on the notion of curvature function sketched in section \ref{pt4}.  

Another physically important abelian system is the electromagnetic field. In this case the action 
cannot be extended by continuity to the class of singular connections that we use here.%
\footnote{
It can be regularized assuming that the field strength is homogeneous inside the $d$-simplices of $({\rm Sd}|\Delta |, \phi )$. The resulting action is the one given by Sorkin \cite{Sorkin:1975jz} for the triangulation $({\rm Sd}|\Delta |, \phi )$. 
}

We conclude this section with the proof of Theorem \ref{CflatAsLimits}. 

{\em Proof of Theorem \ref{CflatAsLimits}.}\\
We have a smooth connection modulo gauge $[A_0] \in {\cal A/G}_{\star , \pi}$ 
which gives us the data of its holonomy and curvature evaluations for every 
$l \in {\cal L}(\star, L)$ and every $S \in {\cal S}(L)$. The $\Delta$-flatness condition 
let us extend the data to define data for the evaluation of holonomy and curvature functions 
for every loop and surface in ${\cal L}(\star, M)_{PL}$ and ${\cal S}(M)_{PL}$ respectively. 
Below we will show that this data in fact determine a $\Delta$-flat connection modulo gauge. 
%
%
According to definition \ref{DeltaFlat} we only need to prove that 
the resulting data induce 
an accumulation point of 
connections in $({\cal A/G}_{\star, \pi}^{\Delta\mbox{-}\infty} )|_{\pi^-}$ in the sense of 
definition \ref{DeltaFlat}. 
The $\Delta$-flatness condition is satisfied by construction.

%

To prove the desired convergence 
we will construct a one parameter family of connections modulo gauge 
\[
[A'_t ] = f_t^\ast [A_0 ] \in ({\cal A/G}_{\star, \pi}^{\Delta\mbox{-}\infty} )|_{\pi^-}
\]
using a one parameter family $f_t : M \to M$ of $\Delta$-smooth homeomorphisms%
\footnote{
A homeomorphism $f : M \to M$ is considered $\Delta$-smooth if it preserves the simplicial decomposition $(|\Delta |, \phi)$ and 
for every $\tau \in (|\Delta |, \phi)$ the function 
$f |_{{\rm Int}(\tau)}: {\rm Int}(\tau) \to f({\rm Int}(\tau))$ is a diffeomorphism. 
} 
with $t \in [0, \infty)$  with the property that for every $l \in {\cal L}(\star, M)_{\rm PL}$ and every $S \in {\cal S}(M)_{\rm PL}$ 
\[
\lim_{t \to \infty} f_t (l) = l_1 \mbox{ and } \lim_{t \to \infty} f_t (S) = S_1  ;
\]
where we recall 
that $l_1$ is our notation for the simplicial representative of $l$, and the analogous notation is used for surfaces (see definition \ref{triangulations}). 
The convergence of loops stated above is in the sense required for the convergence of their associated holonomy functions; any piecewise linear loop is deformed by means of a family of piecewise linear loops whose ``corners" converge to the ``corners" of $l_1$ \cite{Barrett:1991aj}. 
This will imply that for every $l \in {\cal L}(\star, M)_{\rm PL}$ 
\[
\lim_{t \to \infty}{\rm Hol}_l(A'_t) = 
\lim_{t \to \infty}{\rm Hol}_{f_t (l)}(A_0) =
{\rm Hol}_{l_1}(A_0) = {\rm Hol}_{l_1}(A_\Delta) , 
\]
and similarly for every $S \in {\cal S}(M)_{\rm PL}$ 
\[
\lim_{t \to \infty}F_S(A'_t) = 
\lim_{t \to \infty}F_{f_t (S)}(A_0) = 
F_{S_1}(A_0) = F_{S_1}(A_\Delta) . 
\]
According to the notion of convergence stated above, this will 
exhibit 
the connection $[A]_\Delta$ defined by the extended data of holonomy and curvature evaluation, as $[A]_\Delta \in {\cal A/G}_{\star , \pi}^{\Delta\mbox{-flat}}$. 

The rest of the proof deals with the existence of  one parameter family of the 
$\Delta$-smooth homeomorphisms $f_t : M \to M$ with $t \in [0, \infty)$ 
whose limit action in any loop or surface coincides with its simplicial representative as stated above. 

Before we describe the $\Delta$-smooth homeomorphisms we recall some necessary definitions; for a detailed exposition see for example \cite{fritsch-piccinini}. 
$|{\rm Sd}\Delta |$ denotes the geometric realization of the simplicial complex ${\rm Sd}\Delta$; it 
is a simplicial manifold realized as a subset of an euclidean space of large dimension. Since the vertex set of ${\rm Sd}\Delta$ 
is $\Delta$; then $|{\rm Sd}\Delta | \subset \R^\Delta$. 
By definition, the geometric realization of 
each vertex (0 dimensional simplex) $\sigma \in {\rm Sd}\Delta$ is 
$| \sigma |= e_\sigma \in \R^\Delta$; in other words, all the components of the vertex $| \sigma | \in |{\rm Sd}\Delta | \subset \R^\Delta$ 
are zero except for the one corresponding to $\sigma \in \Delta$ which is equal to one. 
Given an abstract simplex $\tau \in {\rm Sd}\Delta$ its geometric realization $|\tau | \subset |{\rm Sd}\Delta | \subset \R^\Delta$ 
is the geometric simplex determined by the geometric realization of its vertices. 
Notice that $|{\rm Sd}\Delta |$ is contained in the subset of $\R^\Delta$ composed by vectors with positive components, 
$|{\rm Sd}\Delta | \subset \R^{+ \Delta} \subset \R^\Delta$. 

Now we define the first ingredient of the construction of our $\Delta$-smooth homeomorphisms. 
We will denote the elements of $\R^\Delta$ as 
$x = \sum x^\sigma e_\sigma \in \R^\Delta$. 
Let $P: \R^\Delta \to \R^\Delta$ be defined by 
$P(x) = \frac{x}{\sum_{\sigma \in \Delta} x^\sigma}$. 
Every point $x \in |{\rm Sd}\Delta | \subset \R^\Delta$ defines a ray though the origin; the map $P$ sends the whole ray to $x \in |{\rm Sd}\Delta |$. 

Now we will construct a one parameter family of maps 
$\tilde{h}_t : |{\rm Sd}\Delta |  \to |{\rm Sd}\Delta |$. 
\[
\tilde{h}_t = P \circ \hat{h}_t \quad , \mbox{ where } 
\hat{h}_t(\sum x^\sigma e_\sigma)= 
\sum e^{t {\rm dim} \sigma}  x^\sigma 
e_\sigma .
\]
It is important to notice that $\tilde{h}_t$ is a homeomorphism which 
preserves simplices; in particular, the vertices are fixed points. 
In addition, for every simplex of dimension bigger or equal to one the restriction of the map to the interior of the simplex, 
$\tilde{h}_t : {\rm Int}(|\tau |) \to {\rm Int}(|\tau |)$, is a diffeomorphism. 

The geometry of $\tilde{h}_t$ is rather simple. First notice that $\hat{h}_t $ is linear, and second recall the property of $P$ sending rays through the origin to the point of that ray that intersects the hyperplane that contains $|{\rm Sd}\Delta | \subset \R^\Delta$. Then the intersection of a linear subspace of $\R^\Delta$ with $|{\rm Sd}\Delta |$ is sent to the intersection of another linear subspace with $|{\rm Sd}\Delta |$ . Thus, according to the euclidean structure of a simplex $\tilde{h}_t$ sends affine hyperplanes to affine hyperplanes; in particular, it sends straight lines to straight lines. 

Now lets see what happens as the parameter $t$ increases. Consider 
$x \in {\rm Int}(|\tau |) \subset  |{\rm Sd}\Delta |$, where $\sigma' \in \Delta$ is the higher dimensional simplex of the vertices of $|\tau |$. It is easy to see that 
\[
\lim_{t \to \infty} (\tilde{h}_t x)^{\sigma'} = 1 , 
\]
while the other components tend to zero. 

Then the image of a a piecewise linear curve $c$ or surface $S$ under 
$f_t = \phi \circ \tilde{h}_t \circ \phi^{-1}: M \to M$ 
is another piecewise linear curve or surface, and in the limit as $t \to \infty$ they approach their simplicial  representatives 
\[
\lim_{t \to \infty} f_t c = c_1 \mbox{ and } \lim_{t \to \infty} f_t S = S_1 . 
\quad \quad
\Box
\]


\section{Family of increasingly finer scales}
\label{pt7}
%
In section \ref{pt5}
we presented a family of 
functions of the connection 
constructed using a simplicial decomposition 
$( | \Delta | , \phi )$ of the smooth manifold $M$.  In fact, since we assumed $M= S^d$ 
we could use a family consisting exclusively of 
curvature functions. 
In order to focus attention we will maintain this assumption. 
A scale is defined by the family of functions available to describe the connection modulo gauge. 
Thus, 
the scale that we considered was defined by the family of 
curvature functions of surfaces $S \in {\cal S}(L)$ that fit in 
$\phi ({\rm Sd}|\Delta |^{(2)})$ with $(|\Delta |, \phi )$ being a triangulation of $S^d$. The corresponding space of curvature evaluations was called 
$\check{\Omega}_L$. 
In this section we will consider a sequence of simplicial decompositions of $S^d$ generated by refining the original triangulation 
$({\rm Sd}^n|\Delta |, \phi )$; see definition \ref{triangulations}. The measuring scales will be labeled by the integer $n$. At scale $n$ 
the original triangulation is refined by baricentric subdivision $n$ times, and 
we consider the family of 
curvature functions of surfaces that fit in 
$\phi ({\rm Sd}^{n+1}|\Delta |^{(2)})$. The family of surfaces will be denoted by ${\cal S}(n)$  (with ${\cal S}(0) = {\cal S}(L)$). 
The plaquettes at scale $n$ are denoted by 
$\sigma \in ({\rm Sd}^{n+1}|\Delta |^2 , \phi)$, 
and the corresponding space of curvature evaluations is called $\check{\Omega}_n$ (or 
$\Omega_{n, \pi}$). 
There are other families of increasingly finer scales; here we just present this one because its construction is simple. 
The process of baricentric subdivision creates finer triangulations, which in some sense become degenerate. 
If we rescale the simplices to have unit $d$ volume their shape can look arbitrarily elongated as the scale gets finer. 
This feature becomes inconvenient to prove convergence of regularized functions. We will comment on this later in this section. 

The reconstruction theorem \ref{ReconstructionContinuum} tells us 
that the family of holonomy functions can separate points in the space of connections modulo gauge. 
Since curvature functions cover holonomy functions when the base is $S^d$,  
the theorem implies that 
given any two different connections modulo gauge 
$[A] , [A'] \in {\cal A/G}_{\star , \pi}$ there is 
a surface $S \in {\cal S}(M)_{\rm PL}$ such that 
$F_S (A) \neq F_S (A')$. 
We are interested in characterizing the connection modulo gauge using the family of curvature functions corresponding to the 
family of surfaces $\cup_n {\cal S}(n)$. 
The following properties of triangulations generated by successive baricentric subdivision will let us reach a concrete statement in this regard: \\
(i) Given any open set $U \subset S^d$ there is an integer $n$ 
such that $U \cap \phi({\rm Sd}^{n+1}|\Delta |^0)$ contains a point 
$p$ such that all the plaquettes 
$\sigma \in ({\rm Sd}^{n+1}|\Delta |^2 , \phi)$ 
touching $p$ are contained in $U$. \\
(ii) For any point $p \in \phi({\rm Sd}^{n+1}|\Delta |^0)$ the set of tangent vectors to the edges in 
$({\rm Sd}^{n+1}|\Delta |^1 , \phi)$ starting at $p$ spans $T_pS^d$. \\
Now, the smoothness of the connection together with properties (i) and (ii) makes the 
the proof of the theorem below immediate. 
\begin{theorem}
\label{SeparatesPoints}
Given any two different connections modulo gauge $[A] , [A'] \in {\cal A/G}_{\star , \pi}$ there is a sufficiently large integer $n$ and a $2$-simplex 
$\sigma \in ({\rm Sd}^{n+1}|\Delta |^2, \phi)$ such that 
\[
F_\sigma (A) \neq F_\sigma (A') . 
\]
\end{theorem}

It is also interesting to study the behavior of regularization as the scale is refined. This will depend on the function that is being regularized. Below we state the convergence result for curvature evaluations $F_S$ for surfaces $S \in {\cal S}(M)_{\rm PL}$ that do not necessarily fit inside 
$\phi ({\rm Sd}^{n+1}|\Delta |^{(2)})$ for any integer $n$. 
In order to simplify notation the 
$L$-curvature evaluation at scale $n$ 
that corresponds to $[A] \in {\cal A/G}_{\star , \pi}$
will be denoted by 
$\check{\omega}_n(A) \in \check{\Omega}_n$. 
\begin{theorem}
\label{ConvergenceOfRegu}
If $\dim M = 2$, 
given any $S \in {\cal S}(M)_{\rm PL}$ and any 
$[A] \in {\cal A/G}_{\star , \pi}$ 
\[
\lim_{n \to \infty} = F^{{\rm reg},n}_S(\check{\omega}_n(A)) = F_S(A) . 
\]
\end{theorem}
{\em Proof.} 
Due to Corollary \ref{regF} $F^{{\rm reg},n}_S : \check{\Omega}_n \to \R$ 
is well-defined and its value at $\check{\omega}_n(A)$ is 
$F_{S_n}(A)$. Thus, our job is to prove the pointwise convergence of a family of functions of smooth connections. 

In the two dimensional case 
\[
F_S(A) - F_{S_n}(A) = F_{D_n}(A) , 
\]
where $D_n = S - S_n \in {\cal S}(M)_{\rm PL}$ as 2-chains. We will prove that the area of $D_n$ according to the euclidean metric of $|\Delta|$ goes to zero as $n$ goes to infinity; this will imply that $F_{S_n}$ converges to $F_S$ pointwise in the space of smooth connections. 
Now we establish a bound for the area of $D_n$. 
\[
{\rm Area}(\phi^{-1}D_n) \leq {\rm Length} (\phi^{-1}\partial S) {\rm Diameter}(n) , 
\]
where ${\rm Diameter}(n)$ is the maximum diameter of all the simplices in 
${\rm Sd}|\Delta | \subset |\Delta |$. Since 
\[
\lim_{n \to \infty}{\rm Diameter}(n) = 0 , 
\]
we can conclude that in the two dimensional case the theorem holds. $\Box$

In higher dimensional cases the problem becomes much more subtle for two reasons: The first reason is that the relative position of the triangulation 
$(|\Delta | , \phi)$ and the surface is not as constrained as in the two dimensional case. The second reason is merely due to our refinement process; the iteration of  baricentric subdivision creates simplicial complexes whose simplices are not only smaller, but whose shapes 
differ more and more from being equilateral. 
Then, it becomes difficult to establish convenient estimates for the area of the simplices in $D_n$ relative to their total number. 
We do not know if the theorem stated above is valid for arbitrary dimension when our method of refinement is baricentric subdivision. However, we know that there are other methods of refinement for which the theorem holds. A description of this result will be given in a future publication.

\subsection{Coarse graining}
%
If we evaluate all the available functions at 
a given measuring scale, we should have enough information to deduce the 
value that observables at a coarser scale would take. 
In the context of the families of curvature functions that we have described the following statement is clear. 
\begin{theorem}
Consider a surface $S \in {\cal S}(n)$ such that $S = S_1 \cup S_2$ for two surfaces $S_1, S_2 \in {\cal S}(n+1)$ with non overlapping 
interiors. Then for any $[A] \in {\cal A/G}_{\star , \pi}$ 
\[
F_S(A) = F_{S_1}(A) + F_{S_2}(A) . 
\]
\end{theorem}

\begin{remark}
\label{FvsH}
There are pairs of connections modulo gauge $[A], [A'] \in {\cal A/G}_{\star , \pi}$ 
and surfaces $S, S_1, S_2$ such that 
\begin{itemize}
\item $S = S_1 \cup S_2$ , $S_1 \cap S_2 \subset \partial S_1$, $S_1 \cap S_2 \subset \partial S_2$, 
\item $F_S(A) \neq F_S(A')$, 
\item ${\rm Hol}_{\partial S_1}(A) \neq {\rm Hol}_{\partial S_1}(A')$, 
${\rm Hol}_{\partial S_2}(A) \neq {\rm Hol}_{\partial S_2}(A')$, 
\item but ${\rm Hol}_{\partial S}(A) = {\rm Hol}_{\partial S}(A')$. 
\end{itemize}
\end{remark}
Examples of such connections are easy to find. The ultimate reason is that there are real numbers $f_1, f_2; f'_1, f'_2$ such that $f_1 + f_2 \neq f'_1 + f'_2$ and 
$\exp{i(f_1 + f_2)} = \exp{i(f'_1 + f'_2)}$.

\section{Lifted parallel transport}
\label{pt8}
%
Here we develop the same theme introduced previously in this paper, but we work 
on ${\cal A}_\pi$, not on ${\cal A/G}_{\star , \pi}$. We will use a given collection of local trivializations and the corresponding transition functions. 
Inside 
${\cal A}_\pi$ the basic type of function 
studied in this section 
is an extended notion parallel transport along open paths defined with respect to a local trivialization. 

Again the theme is that we have partial knowledge about the bundle structure and the connection. Given each d-simplex $\tau_l \in (| \Delta |^d , \phi )$, the bundle $\pi^{-1}(\tau_l)$ is trivial, but at the staring point we do not assume to know a given trivialization of it completely. 
Our partial knowledge contains only the trivialization over the discrete set of points 
$v_k \in (Sd | \Delta |^0 , \phi ) \cap \tau_l$ 
\[
\varphi_l (v_k , \cdot ) : U(1) \to \pi^{-1}(v_k) .
\]
For each vertex contained in two d simplices, $v_k \in \tau_l \cap \tau_m$,  
we know the evaluation of the 
transition function 
$T^{lm}= \varphi_m^{-1} \circ \varphi_l$ at $v_k$. 

The collection of paths considered is that of the edges of the baricentric subdivision, 
$e_k \in ( Sd |\Delta |^1 , \phi )$. The functions 
that measure the connection are the lifted parallel transport functions or 
``logarithm of parallel transport functions'' along those edges 
\[
\tilde{\rm P}_{e_k , l}(A) = \int_{e_k } A ,  
\]
where $A$ in the right-hand side of the equation is the $\R$ valued $1$-form corresponding to the connection in the trivialization of $\pi^{-1}(\tau_l)$, where 
$e_k \subset \tau_l$. Thus, the collection of lifted parallel transport evaluations 
\[
\check{A}_L(A)= \{\tilde{\rm P}_{e_k , l}(A) \in {\rm Lie}(U(1)) = \R\} \in \check{\cal A}_L^\varphi
\] 
is part of our knowledge. 
Since its meaning is relative to our partial knowledge of the local trivialization we include the symbol $\varphi$ in the notation. 
Together with the partial knowledge about the local trivializations described above, they constitute the data available to us at the $L$ ``measuring scale.'' 

The integral that defines $\tilde{\rm P}_{e_k , l}(A)$ exists for every $A \in {\cal A}_\pi$ because $e_k$ is a smooth closed curve (which actually means that it is the restriction of a smooth curve $e_k' \supset e_k$ with an open domain). For non abelian groups the definition of $\tilde{\rm P}_{e_k , l}(A)$ follows the same idea as the one described for the curvature function in section \ref{pt4}; thus, we construct a curve in $G$ by the parallel transport along a curve according to the local trivialization, and we lift the curve to $Lie(G)$ using the origin as starting point. 

This notion of lifted parallel transport lets us recover ordinary parallel transport by exponentiation. Let $c \subset ( Sd |\Delta |^{(1)} , \phi )$ be an oriented curve. To calculate the corresponding parallel transport map we write the curve as a composition of edges, $c = e_n \circ \ldots \circ e_1$, and consider each edge inside one local trivialization $e_i \subset \tau_i$. 
The formula for parallel transport 
${\rm P}_c(A) : \pi^{-1}(c(0)) \to \pi^{-1}(c(1))$ is 
\[
\varphi_n(e_n(1)) 
\circ 
\ldots 
\circ  \exp{(\tilde{\rm P}_{e_2 , 2}(A) )} 
\circ  T^{12}(e_1(1)) 
\circ \exp{(\tilde{\rm P}_{e_1 , 1}(A) )} 
\circ \varphi_1^{-1}(e_1(0)) .  
\]

Also the curvature evaluation of a surface with boundary can be calculated in terms of the data 
$\check{A}_L(A) \in \check{\cal A}_L^\varphi$. Let $S \subset \phi(Sd | \Delta |^{(2)})$ be a surface with boundary. We consider it as a 2 chain and write it as a sum of elementary plaquettes $\sigma_i \in (Sd | \Delta |^{2} , \phi )$, 
$S = \sigma_1 + \ldots + \sigma_n$. 
Therefore,  $F_S(A) = \sum_{i = 1}^{i=n} F_{\sigma_i}(A)$, 
where each of the $F_{\sigma_i}(A)\in \R$ is independent of the local trivialization which can be used to calculate it. For an elementary plaquette 
$\sigma_i$ contained in the d simplex $\tau_l$ 
whose boundary is $\partial \sigma_i = e_{i, 1} + e_{i, 2} + e_{i, 3}$, the curvature evaluation is 
\[
F_{\sigma_i}(A)= 
\tilde{\rm P}_{e_{i, 1} , l}(A) + \tilde{\rm P}_{e_{i, 2} , l}(A) + \tilde{\rm P}_{e_{i, 3} , l}(A) . 
\]
This formula provides a co-boundary map $\check{\cal A}_L^\varphi \to \check{\Omega}_L$. 

In this non gauge invariant variant of the ideas presented in previous sections, the primary functions measuring the connection are the lifted parallel transport functions $\{ \tilde{\rm P}_{e_k , l} \}$. They provide an array of geometrically meaningful averages of the connection, where as the functions $\{ F_\sigma \}$ provide an array of geometrically meaningful averages of the curvature. Moreover, the integrated versions of connection and curvature are related in a natural way. 

Since all the information present in $\check{\omega}_L(A) \in \check{\Omega}_L$ 
is contained in $\check{A}_L(A) \in \check{\cal A}_L^\varphi$, the results presented 
in the previous sections 
(Theorems \ref{CflatAsLimits} and \ref{CurvEvalDetermsDeltaFlat})
characterizing the bundle structure and constraining the connection modulo gauge 
from the data $\check{\omega}_L(A)$, can also be derived from the data $\check{A}_L(A)$. Also the results can be derived again following parallel arguments. We will not show these results here; we will limit ourselves to state the characterization theorem. 
\begin{theorem}
\label{pReconstr}
Let $(E, \pi , M)$ be a smooth principal $U(1)$-fiber bundle with a smooth connection denoted by $A$. 
Among the $U(1)$-fiber bundles with base space $S^d$, 
the bundle structure is characterized (up to bundle equivalence) by the following data defined with the aid of a triangulation $(|\Delta| , \phi)$ of the base space: \\
(i) The evaluation of the local trivialization on the discrete set of points 
$v_k \in (Sd | \Delta |^0 , \phi )$ 
\[
\varphi_l (v_k , \cdot ) : U(1) \to \pi^{-1}(v_k) ,
\]
(ii) The evaluation of the lifted parallel transport along the edges $e_k \in (Sd | \Delta |^1 , \phi )$ 
\[
\check{A}_L(A)= \{\tilde{\rm P}_{e_k , l}(A) \in {\rm Lie}(U(1)) = \R\} \in \check{\cal A}_L^\varphi . 
\] 
In addition, all the connections which induce data $(i)$, $(ii)$ can be 
transformed into 
the same 
$\Delta$-flat connection (modulo gauge) 
$[A_\Delta] \in {\cal A/G}_{\star , \pi}^{\Delta\mbox{-flat}}$ by means of local deformations. 
\end{theorem}


\section{Concluding remarks}

%

We started the paper reminding the reader of a classical differential geometric result behind theories of connections on principal fiber bundles: 
The evaluation of all 
holonomy functions gives enough information to characterize the bundle structure 
(among those sharing the same structure group and base manifold) 
and the connection up to a bundle equivalence map; for a detailed explanation see subsection \ref{pt3.1}. 
We also recalled that there are important approaches to the construction of non perturbative quantum gauge field theories which use discrete sets of holonomy functions as their cornerstone. In this respect, we pointed out that the reconstruction result just described did not apply to these approaches because 
a discrete set of holonomy functions 
does not characterize the bundle and does not constrain the connection modulo gauge appropriately.

An obvious moral can be drawn from the above discussion: considering discrete collections of holonomies as the basic functions that describe the connection at a given macroscopic scale completely changes the nature of 
the gauge field theory. 
The term compactified theory could be used to distinguish the resulting theory from the original which can distinguish between strong and weak fields. 

Our objective was 
to provide a setting to describe effective gauge theories at a given scale. Thus, we studied the problem of 
finding an appropriate  
complement to a discrete set of holonomy functions, providing a better arena for effective gauge theories. 
We showed that in the case of abelian theories, if one complements holonomy data in the way done by the evaluation of their covering function, the curvature function, our partial knowledge of the connection is qualitatively better suited to be the cornerstone for effective abelian theories of connections. 
We exhibited a discrete set of curvature functions, and proved that in the abelian case their 
evaluation characterizes the bundle structure (up to equivalence), and 
constrains the connection modulo gauge up to ``local details'' ignored when working at a given scale; see theorems 
\ref{fReconstr}, \ref{AbelianLocalization} and \ref{CflatAsLimits}. 

For non abelian theories the same issues exist. For example, 
$SU(2)$ bundles over four dimensional base spaces can also be non trivial. 
The non triviality 
of the bundle, and the possibility of not charactering the connection up to a local deformation by the evaluation of holonomy functions, 
shows up in a very similar way to the situation studied in section \ref{pt5}. 
In this paper we do not provide a complete framework to deal with this problem. However, we could say that a discrete set of holonomy functions would need to be complemented by information that keeps track of winding numbers of maps. These winding numbers should characterize homotopy types of holonomy evaluation maps in a way similar to the one shown in our proofs of theorems 
\ref{fReconstr}, \ref{AbelianLocalization} and \ref{CflatAsLimits}. 

Now we conclude with the implications of our results with regard to lattice gauge theory, loop quantum gravity and spin foam models. Our study directly concerns the possible interpretation of a discrete gauge theory on a lattice as an effective theory at the scale given by the lattice itself. 
If the dynamics of the system selects a scale above which large variations of the holonomy function are suppressed and this scale is much smaller than the lattice scale, one expects that the discrete model on the lattice also predicts that 
the expectation value of holonomy functions be close to the identity and their fluctuations be small. 
When this hypothesis holds, our results indicate that 
one could use the lattice theory as the basis of an effective theory provided that: 
(i) the bundle is treated as part of the background, (ii) a set of macroscopic variables {\em different from holonomy functions} is identified. This is needed in order to handle connections whose parallel transport along long distances (measured in lattice spacing units) is not close to the identity, which is essential if we want to have a notion of coarse graining 
that does not identify weak and strong field configurations. 
If we follow this route, the issue of whether the macroscopic behavior of the model is approximately described by the classical theory that was quantized is still very far from obvious. There are many numerical results in the lattice indicating that a central player in this issue is, precisely, the measure of the set of monopole configurations. 

\section*{Acknowledgements}
This work was partially supported by CONACyT grant 80118.


\end{document}